\input{epsf}
\documentstyle[12pt]{article}
\begin{document}
\title{Consistent Quantum Counterfactuals}
\author{Robert B. Griffiths\thanks{Electronic address: rgrif@cmu.edu}\\
Department of Physics\\
Carnegie Mellon University\\
Pittsburgh, PA 15213}

\maketitle

\begin{abstract}
	An analysis using classical stochastic processes is used to construct a
consistent system of quantum counterfactual reasoning.  When applied to a
counterfactual version of Hardy's paradox, it shows that the probabilistic
character of quantum reasoning together with the ``one framework'' rule
prevents a logical contradiction, and there is no evidence within this approach
for any nonlocal influences.  Counterfactual reasoning can support a realistic
interpretation of standard quantum theory (measurements reveal what is actually
there) under appropriate circumstances.
\end{abstract}

	A bullet fired at a beer mug is stopped by a wooden board located
between the gun and the mug.  What {\em would} have happened {\em if} the board
had not been present?  The answer to simple {\it counterfactual} questions of
this sort is intuitively obvious for events in the macroscopic ``classical''
world of everyday experience. Their quantum counterparts, on the other hand,
have given rise to endless controversy.  Suppose a Stern-Gerlach apparatus
measures $S_z=1/2$ for a spin half particle.  {\it Would} $S_z$ have been $1/2$
{\it if} the measurement had not been made?  What {\it would} have been the
result if the apparatus had been set up to measure the spin in a different
direction?  Readers familiar with the EPR paradox and Bell's inequality
\cite{n01,bl87,rd87,de95}, Hardy's paradox \cite{hd92}, and the like are 
probably aware that even admitting that such counterfactual questions might
have answers can be a dangerous first step into a conceptual swamp \cite{n02}.
Nonetheless, counterfactuals seem a necessary part of any realistic version of
quantum theory in which properties of microscopic systems are not simply
``created'' by measurements \cite{de2}.  And if our everyday experiences take
place in a world which is fundamentally quantum mechanical, as most physicists
believe, there must be at least {\it some} cases of valid quantum
counterfactuals having to do with boards and bullets and the like.

	This article will show how to construct a limited system of quantum
counterfactual reasoning able to address the issues mentioned in the preceding
paragraph.  It extends smoothly into the ``classical'' world of everyday
experience, where it gives intuitively sensible answers; alternatively, it
represents a possible way to extend some types of {\it classical}
counterfactual reasoning \cite{n03} into the quantum realm, without fear of
generating contradictions.  It provides a way of understanding a counterfactual
version of Hardy's paradox without coming to the conclusion that quantum theory
involves mysterious nonlocal influences.  And it can serve as one element of a
realistic interpretation of quantum theory.  In its present form it is
restricted to non-relativistic quantum mechanics.

\begin{figure}
\epsfxsize=6.0 in
\epsfbox{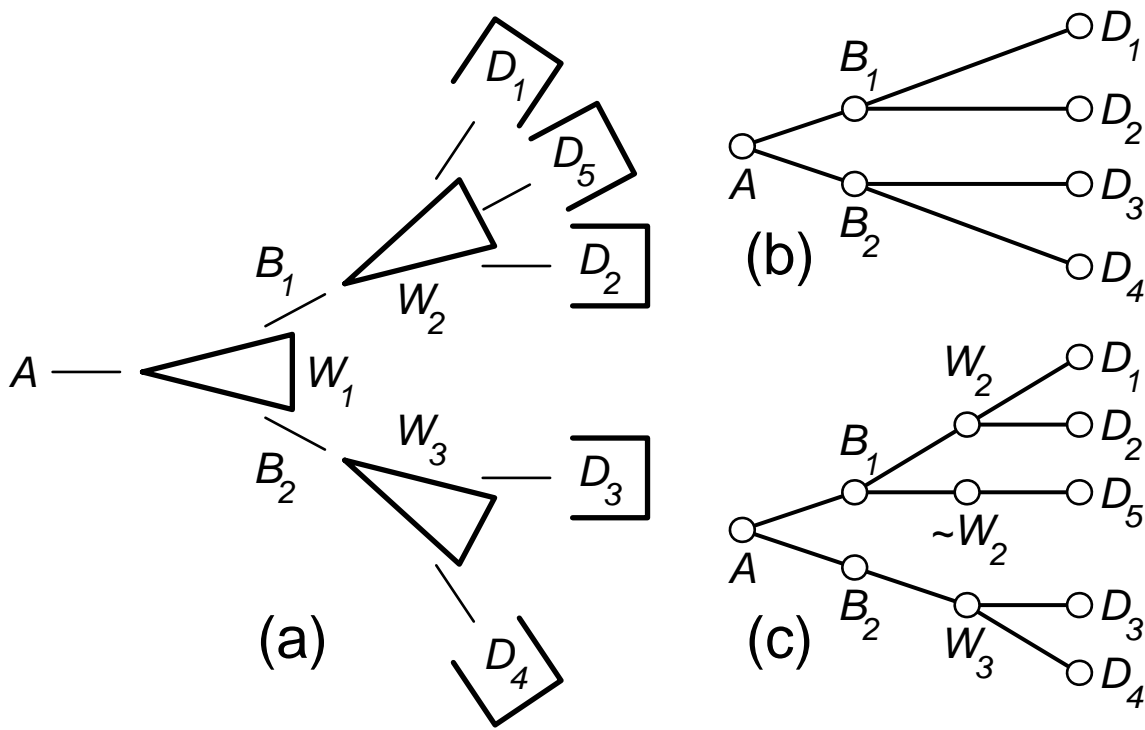}
\caption{%
(a) Particle scattering from wedges. (b) Schematic representation of possible
histories. (c) Histories with wedge $W_2$ present or absent ($\sim\! W_2$).}
\label{fig1}
\end{figure}

	We begin by considering the {\it classical} stochastic process shown in
Fig.~1(a): a particle starting at $A$ moves to the right and scatters randomly
off a set of wedges on its way to a series of detectors, $D_1,D_2\ldots$.
Suppose the particle arrives at $D_1$.  What {\it would have} happened {\it if}
it had suffered a different deflection at wedge $W_2$?  The answer can be read
off of Fig.~1(b), a symbolic representation of possible particle trajectories
or {\it histories}, by starting at node $D_1$, moving backwards in time to the
{\it pivot point} $B_1$, the event immediately preceding the particle's
encounter with $W_2$, and then forward in time along the alternative, or
counterfactual, branch to $D_2$.  Similarly, the question ``What would have
happened had the particle scattered the other way at $W_1$?'' can be answered
by using $A$ as the pivot.  Going forwards in time from $A$ on the alternative
(lower branch) in Fig.~1(b), one sees that two histories, $AB_2D_3$ and
$AB_2D_4$ are possible, and thus the correct answer to this counterfactual
question is not a definite result but instead a {\it probability distribution}
assigning appropriate weights to these alternatives.  Indeed, in a classical
stochastic world, the correct answer to a counterfactual question will in
general be probabilistic; only in special cases will the probability be one.
Even a {\it null counterfactual} question, one for which the antecedent
actually occurred, can have a probabilistic answer.  For example, the answer
to: ``What would have happened had the particle scattered the {\it same} way at
$W_1$?'' (that is, in the same direction in which it actually did scatter in
order to arrive at $D_1$) is a probability distribution with both $D_1$ and
$D_2$ allowed, rather than $D_1$ alone, if we use Fig~1(b) with $A$ as the
pivot.  That null counterfactuals can have this character is typical of a
stochastic, in contrast to a deterministic theory, and has important
applications in the case of quantum counterfactuals, as we shall see.
	The same strategy can be used to address the question ``What would have
happened if wedge $W_2$ had been absent?'' given that $D_1$ was observed.
Assume that a stochastic ``coin flip'', a purely mechanical process with no
human intervention, occurs just before the particle arrives at $W_2$.
Depending on the outcome, the wedge is left in place or yanked out of the way
by a servomechanism.  Using Fig.~1(c), with $B_1$ as the pivot, one obtains
$D_5$ as the answer.  One can also use $A$ as the pivot, in which case the
answer is that either $D_3$, $D_4$, or $D_5$ would have occurred, with certain
probabilities.  Both of these results are intuitively plausible responses to
the counterfactual question, interpreted in slightly different ways.  The fact
that they are different shows the importance of identifying the pivot when one
uses this approach to counterfactual reasoning.

	Our proposal for {\it quantum counterfactual reasoning} is to use {\it
precisely the same method of analysis} applied to a consistent family or
framework of quantum histories: sequences of events represented by orthogonal
projection operators (projectors) to which probabilities are assigned using the
standard dynamical laws of quantum theory \cite{om94,gr96}.  For example,
imagine that the particle in Fig.~1(a) is represented by a wave packet which
scatters off of successive wedges---or think of a photon passing through a
succession of beam splitters.  The events $B_1$ and $B_2$ correspond to
projectors on appropriate regions of space.  A wedge or other scattering center
can be moved out of the way at the last moment based upon a {\it quantum coin
flip}: think of a photon passing through a beamsplitter before triggering one
of two photodetectors connected to suitable amplifiers and a servomechanism.
Counterfactual conclusions are then based upon diagrams of the type shown in
Fig.~1 in the same way as in the classical case.  The main difference between
classical and quantum reasoning comes about through the fact that quantum
events are described using a Hilbert space, and this allows a multiplicity of
stochastic quantum descriptions (frameworks or consistent families or logics)
which are mutually incompatible \cite{om94,gr96}.  Following the usual rules
for consistent histories, we require that a valid counterfactual argument
employ a {\it single} framework; in particular, combining results from two
incompatible quantum frameworks is not allowed \cite{n12}.

	The multiplicity of possible frameworks gives quantum counterfactual
reasoning a slightly different flavor from its classical counterpart, as shown
in the following example.  Imagine that at time $t_0$ a spin half particle with
spin in some direction $w$ is traveling towards a Stern-Gerlach apparatus, and
at $t_1$, shortly before it arrives, a quantum coin is ``flipped''.  Outcome 1
(``heads'') results in a servomechanism orienting the apparatus so that at time
$t_2$ it is in a state $Z$ appropriate for measuring $S_z$, while outcome 2
(``tails'') results in an apparatus state $X$ for measuring $S_x$.  At time
$t_3$ the measurement is complete, and in case 1 (heads) the apparatus is in
one of the two states $Z^+$ or $Z^-$, corresponding to a measurement of
$S_z=\pm 1/2$, while in case 2 (tails) the state is $X^+$ or $X^-$,
corresponding to $S_x=\pm 1/2$.

\begin{figure}
\epsfxsize=5.0 in
\epsfbox{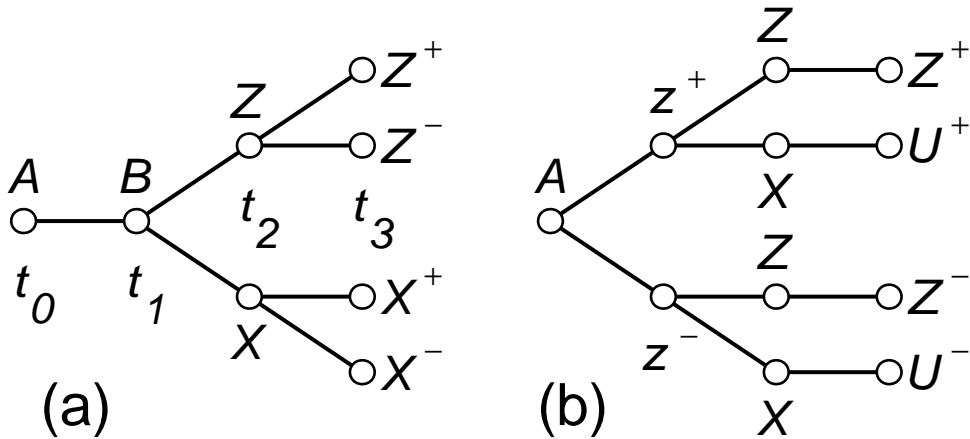}
\caption{%
Two consistent families of histories corresponding to measuring spin components
of a spin half particle in directions determined by flipping a quantum coin.}
\label{fig2}
\end{figure}

	One framework for describing this system is indicated schematically in
Fig.~2(a), where the node $A$ at $t_0$ represents the initial state of the
particle and the apparatus (including the quantum coin), the node $B$ at $t_1$
corresponds to the unitary time development of state $A$, and the nodes at
$t_2$ and $t_3$ indicate the apparatus states before and after the
measurements.  Suppose, for example, that $S_z$ is measured with the result
$Z^+$.  What would have happened if the quantum coin flip had yielded 2,
resulting in a measurement of $S_x$?  The answer, obtained by starting at node
$Z^+$, going backwards to node $B$ (immediately preceding the quantum coin
flip), and then forwards on the lower branch is: $X^+$ or $X^-$ with a certain
probability, depending upon the initial spin direction $w$.  This seems
intuitively plausible.  Applying the same method of reasoning to the null
counterfactual question, ``What would have happened if the quantum coin flip
resulted (as it actually did) in a measurement of $S_z$?'' does {\it not} yield
the answer $Z^+$; instead, there are positive probabilities for both $Z^+$
and $Z^-$ (unless the initial spin direction $w$ coincides with $z$).

	A ``sharper'' answer to this null counterfactual is provided by the
alternative framework in Fig.~2(b) in which the single node $B$ in (a) has been
replaced by two nodes $z^+$ and $z^-$ corresponding to particle spin states
$S_z=\pm1/2$ {\it before} the quantum coin is flipped, and thus before the
measurement.  This framework, which is allowed by consistent histories, though
not by ordinary textbook quantum theory, permits one to say that the
measurement result $Z^+$ reflects the prior state, $z^+$ or $S_z=+1/2$, of the
measured particle \cite{gr96}.  If the node $z^+$ is used as the pivot, the
answer to the null counterfactual question in a case in which $Z^+$ is observed
is that $Z^+$ would have been observed (probability 1) if $S_z$ had been
measured (as it actually was).  However, using this same pivot leads to the
conclusion that had the coin resulted in a measurement of $X$ rather than $Z$,
the result would have been (again with probability 1) the macroscopic quantum
superposition (MQS or Schr\"odinger cat) state $U^+$ which results from a {\it
unitary} time development of a particle with $S_z=+1/2$ interacting with an
apparatus arranged to measure $S_x$.  It is not possible to replace the $U^+$
and $U^-$ nodes in Fig.~2(b) with pairs of $X^+$ and $X^-$ nodes without
violating the standard consistency conditions \cite{gr96}.

	Thus consistent quantum counterfactual reasoning requires the
specification of a consistent family or framework, as does ``ordinary'' quantum
reasoning \cite{gr98}.  Even in a classical stochastic context, English
counterfactual questions are sometimes ambiguous (``What would have happened if
the particle had not arrived at $D_1$?'') because the pivot is not specified,
and in the quantum context the ambiguity can be even worse, because the English
wording (``What would have happened if $S_x$ had been measured rather than
$S_z$?'') can often be modeled by different frameworks, as well as by different
pivots within the same framework \cite{n10}.  The confusion caused by this
ambiguity is illustrated in the following counterfactual version \cite{mr94,st97}
of Hardy's paradox.

	Imagine two spin half particles in an appropriate entangled spin state
sent in opposite directions to two distant detectors, one on the left and one
on the right.  Each detector has two switch settings, 1 and 2, which determine
which spin component will be measured, and the setting is determined by the
flip of a quantum coin, incorporated into the detector, just before the arrival
of the corresponding particle.  A red light on the detector flashes if the spin
component in the measured direction is positive, and a green light if it is
negative.  Each run yields a result such as 1G$\cdot$2R, meaning that the left
detector had a switch setting of 1 and flashed green, and the right detector
with a switch setting of 2 flashed red.  The entangled particle state 
\cite{mr94} is such that 1G$\cdot$2G, 2G$\cdot$1G and 1R$\cdot$1R cannot 
occur (zero probability); all other possibilities occur with finite
probability.

	To construct a paradox, consider a run in which 2G$\cdot$2G is observed
and ask what would have happened {\it in this particular experimental run} if
the quantum coin associated with the right detector had produced a switch
setting of 1 rather than 2.  In the absence of long-range influences (in a
gedanken experiment, detectors can be arbitrarily far apart!), a last minute
change in the switch setting on the right cannot influence the detector on the
left, which would, therefore, have flashed green.  But since 2G$\cdot$1G never
occurs, we conclude that if the switch on the right had been 1, the result in
this run would have been 2G$\cdot$1R.  Applying the same argument with the
roles of the two detectors interchanged, we conclude that if the switches had
been 1 and 2 on left and right, the result would have been 1R$\cdot$2G.  But
then, continuing the argument, had both switches been 1, the result would have
been 1R$\cdot$1R, contradicting the fact that 1R$\cdot$1R never occurs.
Contradictions of this sort easily gives rise to the idea that quantum
mechanics involves some mysterious nonlocality: changing the switch setting on
the right really does influence the particle or the apparatus on the left in
some way.

	The basic counterfactual question which must be addressed in thinking
about this paradox is the following: given a case in which 2G$\cdot$2G is
observed, what would have occurred had {\it both} of the switch settings been
equal to 1?  In order to study it using the system of counterfactual reasoning
introduced above, one needs to find a pivot at a time which precedes both coin
flips, since one is comparing cases in which both of them turned out to be the
reverse of what actually occurred.  This pivot must involve a quantum state of
the two-particle system, since at this earlier time the apparatus states are
not playing a significant role.  The successive counterfactual steps in the
preceding paragraph can then be thought of as devices for finding such a pivot.
Viewed in this way, one finds that even the first step of the argument is
problematical.  In order to infer 2G$\cdot$1R from 2G$\cdot$2G, one must use an
earlier state of the two particles which provides a null counterfactual for the
left detector: that is, which leads with probability one to the conclusion that
its light would (certainly) have flashed green.  Simply using the wave function
of the original two-particle state as it develops unitarily in time will not
support such a sharp conclusion, as there is then a finite probability of 2R
for the left detector \cite{n11}; see the previous discussion of Fig.~2(a).  A
framework in which at an earlier time the left particle is in one of the two
states corresponding to the results 2G and 2R of a later measurement, analogous
to the $z^+$ and $z^-$ states of Fig.~2(b), will yield the desired null
counterfactual for the left detector. But there is a price to be paid: when the
switch on the left detector is changed from 2 to 1, quantum consistency
requires the use of an MQS state for the left detector---see the preceding
discussion associated with Fig.~2(b)---which will obviously prevent completing
the full counterfactual argument of what would have happened had both switches
been 1.  Note that the MQS problem arises in the course of constructing a
correct quantum (counterfactual) description of a {\it single detector}, the
one on the left, {\it not both detectors}.  Nonlocality plays no role: what
matters is consistent quantum counterfactual reasoning applied to a local
measurement.

	The framework just discussed is not unique; there are various
possibilities which can be used to justify different steps in the
counterfactual argument.  However, there is no way to combine these into a {\it
single} framework, as required by the rules of counterfactual reasoning
introduced above, in order to complete the entire counterfactual argument in
which both switches are changed from 2 to 1.  Were there such a framework it
would imply that 1R$\cdot$1R can occur with positive probability, at least as
large as the probability of 2G$\cdot$2G.  But standard quantum mechanics gives
probability zero for 1R$\cdot$1R, as does any framework or consistent family in
which the probability of 1R$\cdot$1R makes sense \cite{gr96n}.  Hence there
cannot be a single framework in which the counterfactual inference leading to a
paradox is valid.  From this perspective, the Hardy paradox can be thought of
as illustrating one of the ways in which classical reasoning can fail when
applied in a quantum context.

	Consistent quantum counterfactuals can generate positive results as
well as block paradoxes.  Consider the well-known EPR-Bohm arrangement in which
two spin-half particles in a singlet state fly apart, and the $x$ component of
the spin of the particle on the right is measured to be $S_x=+1/2$.
Assuming that the particle on the left continues on its trajectory without
interacting with anything, one can infer (probability 1) that it has $S_x=-1/2$  
both before and after the spin measurement on the right \cite{gr87}.  But would
it still have had $S_x=-1/2$ if $S_z$ instead of $S_x$ had been measured for
the particle on the right?  The answer is ``yes'' if one adopts a framework in
which $S_x$ values for the particle on the left make sense at different times
\cite{ndes}.  The argument is straightforward, and only requires combining a
previous consistent history analysis \cite{gr87} of EPR-Bohm (without using
counterfactuals) with the proposal in the present article.  Again, there is no
indication of any mysterious nonlocal influence.

	In addition, counterfactual definitions of quantum properties are
possible under appropriate circumstances.  The analysis given in Sec.~5 of
\cite{gr84} indicates that as long as the behavior of a closed quantum system
is described using a consistent family, the corresponding events can, in
principle, be checked by idealized measurements.  If these are thought of as
carried out with the help of appropriate quantum coins, the procedure for
counterfactual reasoning given above supports a realistic interpretation of the
corresponding quantum events: they would still have occurred even if no
measurements had been made \cite{nreal}.

	In conclusion, the system of consistent quantum counterfactuals
presented here, while of limited scope, is sufficiently powerful to deal with a
number of non-trivial issues in quantum foundations, and could well prove to be
a useful tool for getting rid of some of the ghosts which have long plagued
that discipline.

	Conversations and correspondence with R.~Clifton, P.~Eberhard,
B.~d'Espagnat, N.~D.~Mermin, and H.~P.~Stapp are gratefully acknowledged,
together with comments from various anonymous referees, and financial support
from the National Science Foundation through grant PHY 96-02084.


\begin{thebibliography}{11}
\bibitem{n01}
	There is an enormous and ever increasing literature devoted to this
subject. Various items in \cite{bl87} are extremely helpful.  Critical
analyses and references to the literature will be found in \cite{rd87} and
\cite{de95}
\bibitem{n12}
	This is the quantum counterfactual counterpart to the
``single logic'', p.\ 163 of \cite{om94}, or ``single framework'',
Sec.\ III of \cite{gr96}, rule for ordinary quantum reasoning.
\bibitem{n10}
	If the initial state, pivot, and various later events are all
specified, the probabilities of interest will be independent of which framework
contains these events, for reasons noted in Sec.~5.4 of \cite{om94} and Sec.~IV
of \cite{gr96}.  Note, in particular, the need for an explicit pivot.
\bibitem{n11}
	  Quite generally, if one is limited to the consistent families
employed in textbook versions of quantum theory, for which measurement outcomes
are only related by probabilities to prior states of the measured system,
counterfactual reasoning cannot come to very sharp conclusions. Thus within
this context, Redhead's criticism, Sec.~4.2 of \cite{rd87}, of earlier
counterfactual arguments by Stapp and Eberhard (see \cite{n02}) appears to be
justified.
\bibitem{nreal}
	For some further remarks about quantum reality, and the way in which it
differs from its classical counterpart, see \cite{gr98}.
\end{thebibliography}
\end{document}